\documentstyle[preprint,12pt]{aastex}
\def\arcs{$''$}

\begin{document}
\title{Galaxies at $z\sim7-8$: $z_{850}$-dropouts in the
Hubble Ultra Deep Field}
\author{R.J. Bouwens$^{2}$, R.I. Thompson$^{3}$,
G.D. Illingworth$^{2}$, M. Franx$^{4}$,P. van Dokkum$^{5}$,
X. Fan$^{3}$, M.E. Dickinson$^{6}$, D.J. Eisenstein$^{3}$,
M.J. Rieke$^{3}$}
%        }
\affil{1 Based on observations made with the NASA/ESA Hubble Space
Telescope, which is operated by the Association of Universities for
Research in Astronomy, Inc., under NASA contract NAS 5-26555.}
\affil{2 Astronomy Department, University of California, Santa Cruz,
CA 95064}
\affil{3 Steward Observatory, University of Arizona, Tucson, AZ 85721.}
\affil{4 Leiden Observatory, Postbus 9513, 2300 RA
Leiden, Netherlands.}
\affil{5 Department of Astronomy, Yale University, New Haven, CT 06520.}
\affil{6 National Optical Astronomy Obs., P.O. Box 26732, Tucson, AZ 85726.}

\begin{abstract}

We have detected likely $z\sim7-8$ galaxies in the $144''\times144''$
NICMOS observations of the Hubble Ultra Deep Field.  Objects are
required to be $\geq3\sigma$ detections in both NICMOS bands,
$J_{110}$ and $H_{160}$.  The selection criteria for this sample are
$(z_{850}-J_{110})_{AB}>0.8$, $(z_{850}-J_{110})_{AB} > 0.66
(J_{110}-H_{160})_{AB}+0.8$, $(J_{110}-H_{160})_{AB}<1.2$, and no
detection at $<8500\AA$.  The 5 selected sources have total magnitudes
$H_{160,AB} \sim 27$.  Four of the five sources are quite blue
compared to typical lower--redshift dropout galaxies and are clustered
within a $1\,\sq '$ region.  Because all 5 sources are near the limit
of the NICMOS data, we have carefully evaluated their reality.  Each
of the candidates is visible in different splits of the data and a
median stack.  We analyzed several noise images and estimate the
number of spurious sources to be $1\pm1$.  A search using an
independent reduction of this same data set clearly revealed 3 of the
5 candidates and weakly detected a 4th candidate, suggesting the
contamination could be higher.  For comparison with predictions from
lower redshift samples we take a conservative approach and adopt four
$z\sim7-8$ galaxies as our sample.  With the same detection criteria
on simulated datasets, assuming no-evolution from $z\sim3.8$, we
predict 10 sources at $z\sim7-8$, or 14 if we use a more realistic
$(1+z)^{-1}$ size scaling.  We estimate that the rest-frame continuum
$UV$ ($\sim1800\AA$) luminosity density at $z\sim7.5$ (integrated down
to $0.3L_{z=3} ^{*}$) is just $0.20_{-0.08}^{+0.12}\times$ that found
at $z\sim3.8$ (or $0.20_{-0.12}^{+0.23}\times$ including cosmic
variance).  Effectively this sets an upper limit on the luminosity
density down to $0.3L_{z=3} ^{*}$.  This result is consistent with
significant evolution at the bright end of the luminosity function
from $z\sim7.5$ to $z\sim3.8$.  Even with the lower UV luminosity
density at $z\sim7.5$, it appears that galaxies could still play an
important role in reionization at these redshifts, though definitive
measurements remain to be made.

\end{abstract}


\section{Introduction}

From the spectroscopic identification of a population of $z\sim3$
dropouts (Steidel et al.\ 1996) to recent work on $i$-dropouts (Yan et
al.\ 2003; Stanway et al.\ 2003; Bouwens et al.\ 2003b; Dickinson et
al.\ 2004), the frontier for high redshift galaxy studies is
continually being redefined.  In this paper, we extend this frontier
to $z\sim7$ and beyond by performing a $z_{850}$-dropout search over
the area of the Hubble Ultra Deep Field (Beckwith et al.\ 2004) with
deep NICMOS coverage (Thompson et al.\ 2004a).  The exceptional depth
of both the optical and infrared data makes this area ideal for
carrying out such a search, reaching to 29.5, 29.7, 29.4, 28.8, 27.6,
and 27.4 ($5\sigma$, 0.6\arcsec-diameter apertures) in the $F435W$,
$F606W$, $F775W$, $F850LP$, $F110W$, and $F160W$ bands (hereinafter,
$B_{435}$, $V_{606}$, $i_{775}$, $z_{850}$, $J_{110}$ and $H_{160}$,
respectively.)  Previously, this redshift range had been probed by an
$I_{814}$-dropout search in the HDF-North (Dickinson 2000) and similar
dropout searches around lensing clusters (Kneib et al.\ 2004; Pell{\'
o} et al.\ 2004).  All magnitudes are expressed in the AB system.  We
assume $(\Omega_M,\Omega_{\Lambda},h) = (0.3,0.7,0.7)$ (Bennett et
al.\ 2003).

\section{Analysis}

Our search area was the 0.09$''$ pixel $144''\times 144''$ NICMOS
mosaic (Thompson et al.\ 2004a).  Sources were identified in the
summed $J+H$ image (RT) and the $\chi^2$ (Szalay et al.\ 1999) image
(RB) using the SExtractor code (Bertin \& Arnouts 1996).  Colors were
calculated using a scaled aperture Kron magnitude (1980) with the Kron
factor equal to 1.2.  Total magnitudes were then derived using the
$\chi^2$ image to correct these fluxes to a much larger aperture
(where the Kron factor was equal to 2.5) (see Bouwens et al.\ 2003a).
Typical corrections were $\sim0.8$ mag for each object.
 
\textit{(a) $z_{850}$-dropout selection.}  Objects were required to be
null detections ($<2\sigma$) in the deepest ($V_{606}$ and $i_{775}$)
optical bands (in 0.6\arcsec$\,$-diameter apertures), and lie in the
expected place [$(z_{850}-J_{110})>0.8, (z_{850}-J_{110})_{AB} > 0.66
(J_{110}-H_{160})_{AB} + 0.8$, $(J_{110}-H_{160})_{AB}<1.2$] in the
standard two-colour $z_{850}-J_{110}/J_{110}-H_{160}$ diagram.  To
clean our catalog of possible spurious detections, objects were
required to be $3\sigma$ detections (0.6\arcsec$\,$-diameter aperture)
in both the $J_{110}$ and $H_{160}$ bands.  These procedures
identified a set of 8 sources to a limiting magnitude of
$H_{160,AB}\sim28$.  A separate selection by RT identified a similar
set of objects.  After identification, each source was located in the
original exposures (16 in each band) to ensure that they did not arise
from a small subset of the exposures (e.g., from a pre-integration
cosmic ray hit).  Three of our 8 sources were rejected, being visible
in only a couple of exposures.  The 5 real sources are shown in Figure
1 and Table 1.

Table 1 includes the photometric information for our 5 candidates plus
one red galaxy which nearly met our criteria (this latter object was
also found by Yan \& Windhorst 2004).  Candidates had $H_{160,AB}$
magnitudes ranging from 26.0 to 27.3, or 0.5-1.5 times the
characteristic rest-frame UV luminosity $(L^{*})$ for Lyman break
galaxies at $z\sim3$ (Steidel et al.\ 1999).  Candidates appear to be
rather clustered, with 4 of the 5 candidates falling within a
$\sim1\,\sq '$ area.  Figure 2 displays postage stamp images of each
candidate along with its position in color-color space and an SED fit
to the broadband fluxes.

\begin{figure}[h]
\epsscale{0.95}
\plotone{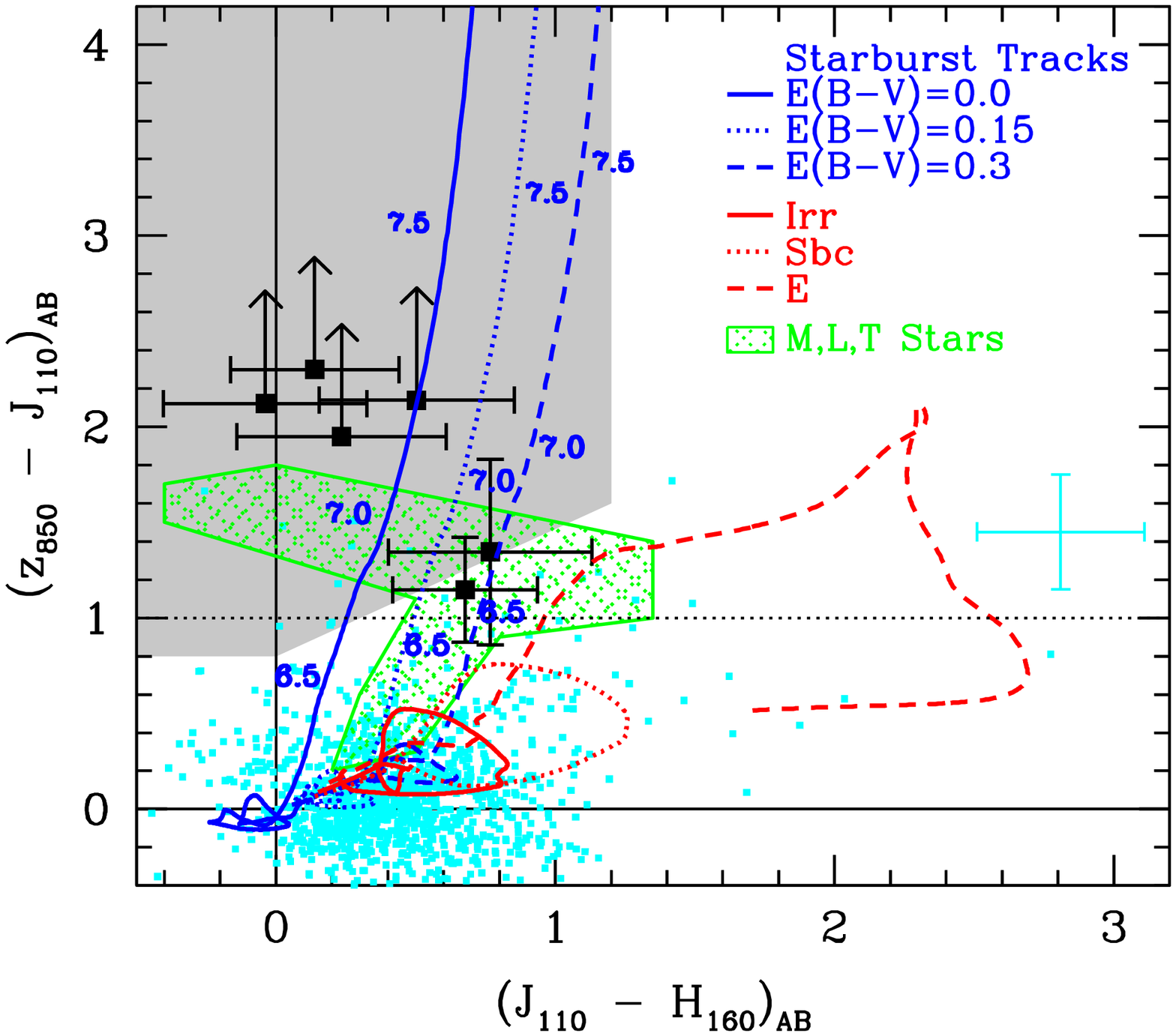}
\caption{$(z_{850}-J_{110})_{AB}/(J_{110}-H_{160})_{AB}$ color-color
diagram showing the position of our $z_{850}$-dropouts (selection
region is shaded gray) relative to the UDF photometric sample (cyan
squares).  Objects included in the source list (Table 1) are shown as
black squares ($2\sigma$ lower limits are indicated by vertical
arrows).  These objects are not detected in the optical $V_{606}$ and
$i_{606}$ bands.  The cyan squares that lie in the selection region
have clear $V_{606}$ and $i_{775}$ detections ($>2\sigma$) and so are
not candidate $z_{850}$-dropouts; representative error bars for these
objects are shown at the right of this diagram.  The color-color
tracks of both lower redshift interlopers (\textit{red lines}) and
$10^8$ yr starburst SEDs with different reddenings (\textit{blue
lines}) are plotted as a function of redshift.  The position of M, L,
and T dwarfs are also shown (\textit{green cross hatched region})
(Knapp et al.\ 2004).  Error bars on the $z_{850}-J_{110}$ and
$J_{110}-H_{160}$ colors are $1\sigma$.}
\end{figure}

\textit{(b) Testing Source Reality.}  Our 5 candidates were then
subjected to several additional tests.  Each source was verified to
exist at the $>2.5\sigma$ level in the $J+H$ image for each of the two
epochs (taken two months apart and at a 90 degree angle to each
other).  Each source was also evident ($>2.4\sigma$) in a median
stacking of the 16 overlapping exposures for each band.  This is
useful since the median process should eliminate sources with flux in
only a few exposures.  After performing the above sanity checks on our
candidates, we repeated our selection procedure on three different
images sets to examine the likelihood that our candidates are simply
spurious detections.  These three images include the ``negative''
images, the first epoch images subtracted from the second epoch
images, and the second epoch images subtracted from the first.  These
images should have similar noise characteristics to the data, but
contain no real sources.  Only 1, 2, and 0 objects, respectively, were
found on each of the above three image sets (5.76 arcmin$^2$) using an
identical selection procedure.  This suggested a small level of
contamination from spurious sources ($1\pm1$ object) in the current
sample.

\textit{(c) An Independent Check on Source Reality.}  An independent
reduction of the NICMOS images was kindly made available to one of us
(RT) by Robberto et al.\ (2004).  The image was inspected by RT and 3
of our 5 candidate sources clearly appear in those images.  However,
no signal is evident at the position of UDF-818-886, while UDF-491-880
is only weakly detected.  Until this is resolved, the contamination may
be higher than estimated above (\S2b).

\begin{figure}
\epsscale{1.0}
\plotone{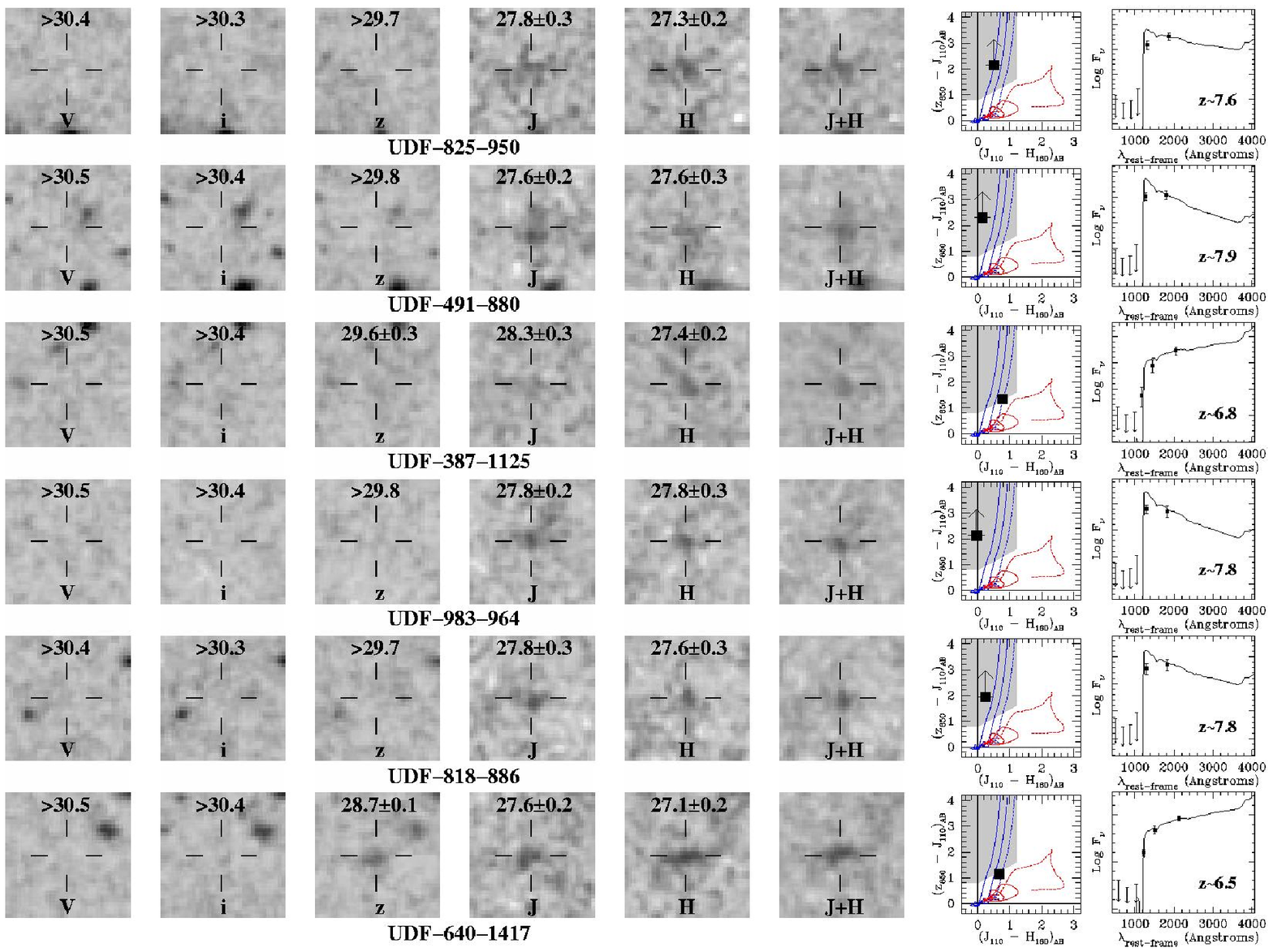}
\caption{Postage stamps images ($V_{606}i_{775}z_{850}J_{110}H_{160}$
bands) of our 5 $z_{850}$-dropout candidates.  Also shown
(UDF-640-1417: bottom row) is one very red
$(z_{850}-J_{110})_{AB}=1.1$ object which nearly met our selection
criteria and could be a reddened starburst at $z\sim6.5$ (or a
reddened early type at $z\sim1.6$) (also found by Yan \& Windhorst
2004).  While our best-fit to UDF-387-1125 is a $z\sim6.8$ starburst
spectrum, this object is also consistent with being a compact
$z\sim1$, $0.01L^{*}$ dust-reddened early type galaxy.  Magnitudes are
those measured in a $0.6''$-diameter aperture.  The three rightmost
panels show the combined $J_{110}+H_{160}$ image for each object, its
position in color-color space, and an SED fit to the broadband fluxes.
The derived redshift is also provided in the rightmost panel.  The ACS
cutouts here are shown at a much higher contrast than the NICMOS
cutouts, demonstrating the significance of the optical non-detections.
The postage stamps are 2.9\arcsec$\times$2.9\arcsec$\,$ in size.  A
linear stretch is used for scaling the pixel fluxes.}
\end{figure}

\textit{(d) Low-Redshift Contamination.}  To test for possible
contamination from low-redshift interlopers, we randomly assigned the
colors of bright ($23.5<H_{160,AB}<25$) objects from the UDF to faint
objects in our field, added photometric scatter, and then repeated our
selection.  No objects were found, suggesting minimal contamination
from low-redshift interlopers.  Possible contamination from T dwarfs
was also considered, given their position in color-color space (Figure
1) and predicted numbers (0.04-0.3 objects) over our field of view
(Burgasser et al.\ 2004).  However, this proved not to be a concern
for our sample, since T dwarfs would appear as $\gtrsim5\sigma$ point
sources in the deep $z_{850}$-band images, and none were found.

\textit{(e) Expected Numbers/Incompleteness Tests.}  It is interesting
to compare the number of candidates against that predicted assuming
no-evolution from lower redshift.  As in other recent work, we adopt a
$z\sim3.8$ $B$-dropout sample from the GOODS fields (Bouwens et al.\
2004, hereinafter B04) as our reference point and project it to
$z\sim6-10$ using our well-established cloning machinery (Bouwens et
al.\ 1998a,b; Bouwens et al.\ 2003a; B04).  Such simulations are
important for establishing the incompleteness, which can be as high as
75\% for these $z\sim7-8$ objects (this includes the effect of
possible blending with foreground galaxies).  Adding the cloned
galaxies directly to the data, we repeat our selection procedure and
thereby derive a no-evolution prediction; this yields 10 dropouts.
However, we know that galaxies evolve in size (a $(1+z)^{-1}$ size
scaling for fixed luminosity: Bouwens et al.\ 2004a,c; Ferguson et
al.\ 2004) and hence surface brightness.  Including this effect, 14
objects are found.  Steeper size scalings (e.g., $(1+z)^{-2}$) yield
still larger values ($\sim18$ objects) while using bluer colors (e.g.,
UV slopes $\beta\sim-2.5$) has little effect on the predictions.

\textit{(f) Source Characteristics / Possible Concerns.} Given the
depth of the UDF $z_{850}$-band imaging, it was somewhat surprising
that only 1 of our 5 candidates is detected in this band.  We used the
simulations described above (\S2e) to quantify this and found that
$58\%$, or 2.9 of our 5 candidates, should be detected at $>2\sigma$
in the $z_{850}$-band.  A single detection in $z_{850}$ has only a
10-22\% likelihood of occurrence, the larger number for significant
clustering.  Four of the objects are spatially clustered, falling
within a $1\,\sq '$ area.  Such clustering is not unexpected, and the
lack of $z_{850}$-band flux would result if they are also at $z>7.5$.

These 4 clustered objects are also quite blue, relative to the
fiducial $10^8$ yr starburst, though the significance of this result
is modest ($<2\sigma$). Their rest-frame UV colors (with $\beta\sim
-3$) are bluer than typical dropouts at both $z\sim3$ (Steidel et al.\
1999) and $z\sim6$ (Stanway et al.\ 2004b) where $\beta\gtrsim-2.2$
(the $E(B-V)\gtrsim0$ track in Figure 1).  If real, they are also
bluer than one would expect from models which give $\beta\gtrsim -2.5$
regardless of age, dust, or metallicity content, including population
III objects (Schaerer 2003; Venkatesan et al.\ 2003).  However, such
blue colors are not completely unknown at $z\sim3$ (e.g., Adelberger
\& Steidel 2000).  They also might arise from a significant
contribution of Ly-$\alpha$ emission (rest-frame EWs $\gtrsim200\AA$)
to the $J_{110}$-band flux.

The one red object in our $z\sim 7-8$ sample has colors that are also
consistent with a $z\sim 1$ galaxy, though it would be rather unusual,
needing to be a compact ($\sim 1$kpc), significantly reddened, old
stellar population dwarf galaxy, $>6$ magnitudes fainter that L$^*$;
we consider it more likely to be at high redshift.

\section{Luminosity Density and Implications}

We are now prepared to compare the observations with the predictions
made earlier (\S2e).  This will permit us to set important constraints
on the evolution at the bright end of the luminosity function in
rest-frame continuum-$UV$ ($\sim1800\AA$) and therefore make
inferences about changes in the luminosity density.  To be
conservative, we shall assume the number of $z_{850}$-dropouts is
four.  Given possible concerns about their validity (\S2b;\S2c;\S2f), we
will also consider the implications if there are even fewer sources.
For the expected number of $z_{850}$-dropouts, we use 14, the
prediction from the $(1+z)^{-1}$ size scaling (\S2e).

Comparing our 4 fiducial candidates with the 14 objects predicted
suggests that the number of objects at the bright end of the LF at
$z\sim7.5$ is just $0.29\times$ that at $z\sim3.8$ (Figure 3).  This
decreases to $0.14\times$ and $<0.13\times$ ($1\sigma$) if only two or
none of our candidates are real, respectively.  Obviously, there are
substantial uncertainties in the estimated shortfall, both as a result
of the small number statistics and the expected cosmic variance
(factor of 2: assuming a CDM power spectrum normalized to high
redshift observations and a redshift selection window of unit width,
e.g., Somerville et al.\ 2004).  Therefore, even no evolution is
consistent with the present result at the $1.5\sigma$ level.

While a number of options are open, the most likely case is a drop of
at least $3.5\times$ in the number of objects at the bright end of the
LF.  Since this is similar to what is found at $z\sim6$ (Stanway et
al.\ 2003; Dickinson et al.\ 2004; Stanway et al.\ 2004a), it is
likely a continuation of the same effect.  A key question is whether
the observed deficit continues all the way down the luminosity
function or if it is due to evolution in the characteristic luminosity
($L^{*}$) at high redshift.  This whole issue is pivotal for questions
about reionization since it is at faint magnitudes that the bulk of
the flux arises (assuming a steep $\lesssim-1.5$ faint-end slope
$\alpha$).  Fortunately, the fainter $i$-dropouts from the UDF are
beginning to provide us with some clues, and some early studies are
already suggesting that the principal form of the evolution is in
luminosity or a steepening of the faint-end slope (Dickinson et al.\
2004; Yan \& Windhorst 2004; Bouwens et al.\ 2004d; cf. Bunker et al.\
2004).  If true, this would provide a natural explanation for our
shortfall and may allow for substantial star formation at higher
redshifts as suggested by recent measurements from WMAP (Kogut et al.\
2003) or the large stellar masses found in the $z\sim6.5$ Kneib et
al.\ (2004) object (Egami et al.\ 2004).  It would also suggest that
for a proper census of these objects the present surveys need to be
extended to considerably fainter magnitudes (with WFC3 and ultimately
with JWST).

In light of the uncertainties regarding the form of the evolution, we
have chosen simply to quote the evolution in luminosity density down
to the total magnitude limit of our survey ($H_{160,AB}\sim27.5$, or
$\sim0.3L_{z=3} ^{*}$).  Conversions to star formation rate density
(uncorrected for extinction) are made using the now canonical
conversion factors for the Salpeter IMF (Madau et al.\ 1998).  For
both quantities (the luminosity density and the star formation rate),
we infer a larger drop than above (due to luminosity-weighting).  To
the faint end limit and including the Poissonian variations quoted
above, we find that $\rho(UV,z=7.5)/\rho(UV,z=3.8) =
0.20_{-0.08}^{+0.12}$ using our fiducial list of candidates and
$0.10_{-0.05}^{+0.09}$ and $<0.05$ ($1\sigma$) assuming only two or
none of our candidates are real, respectively.  Uncertainties on these
quoted factors increase to $0.20_{-0.12}^{+0.23}$,
$0.10_{-0.07}^{+0.20}$, and $<0.11$ ($1\sigma$), respectively,
including the expected field-to-field variations (cosmic variance)
quoted above.  Figure 3 shows a comparison of these results with those
at lower redshift.

This is the first such deep sample ever compiled at $z\sim7-8$ and
allowed us to set some constraints on the bright end of the rest-frame
$UV$-continuum luminosity function at $z\sim7.5$, during the epoch of
reionization.  The similarity of the present result with that at
$z\sim6$ (Stiavelli et al.\ 2004; Yan \& Windhorst 2004) suggests that
galaxies could have been an important contributor to reionization at
these early times, though a characterization of their role warrants
more definitive measurements.

\begin{figure}
\plotone{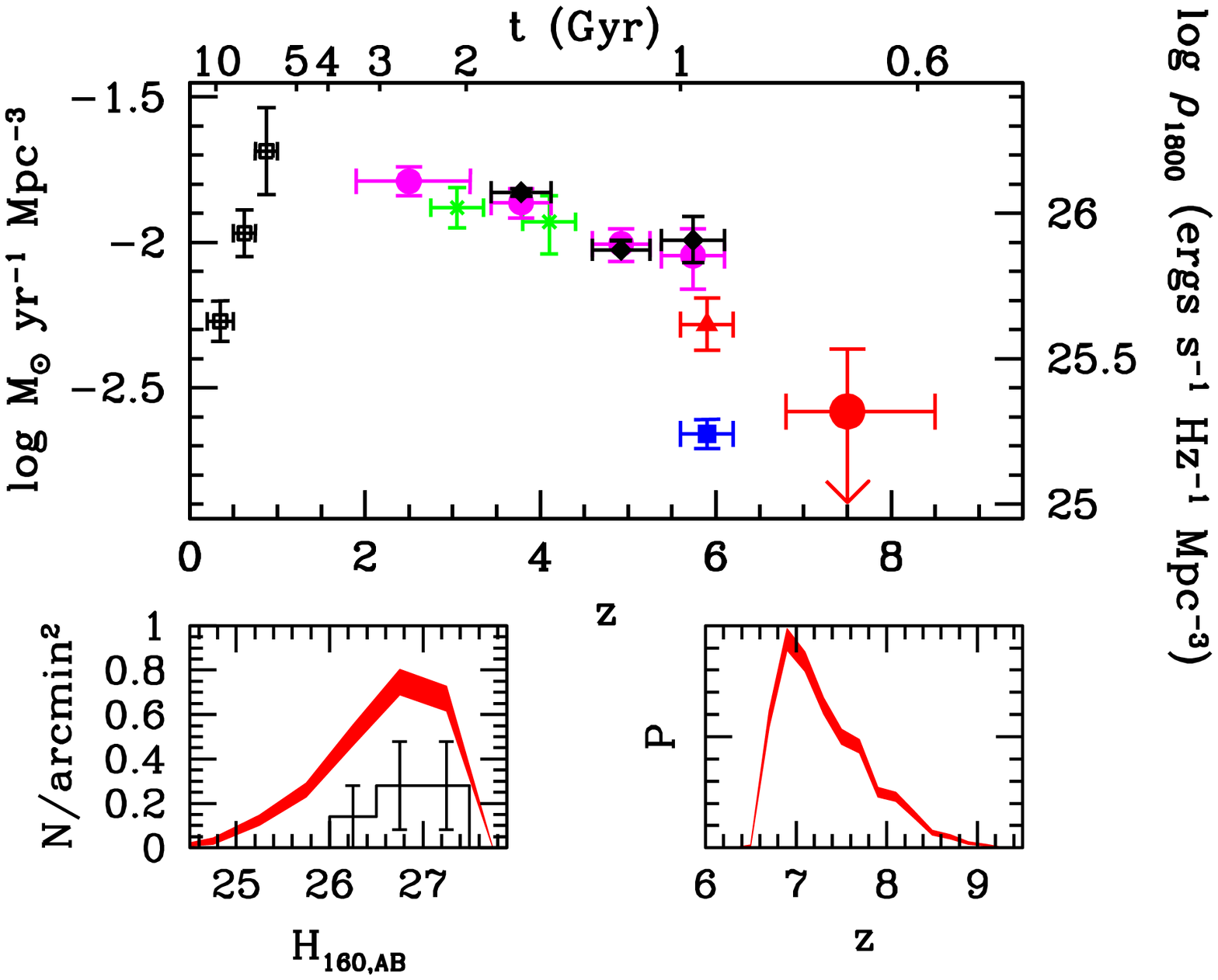}
\caption{Top Panel: Rest-frame continuum $UV$ ($\sim1800\AA$)
luminosity density (integrated down to 0.3 $L_{z=3} ^{*}$)
vs. redshift.  The observed luminosity density is converted to a star
formation rate (uncorrected for extinction) assuming a Salpeter IMF
(e.g., Madau et al.\ 1998).  The present determination (assuming 4
candidates) is shown as the large red circle, with an upper limit
shown to acknowledge possible concerns regarding several of our
candidates.  Previous determinations from Lilly et al.\ (1996)
(\textit{open squares}), Steidel et al.\ (1999) (\textit{green
crosses}), Giavalisco et al.\ (2004a) (\textit{solid black diamonds}),
Bunker et al.\ (2004) (\textit{solid blue square}), B04 (\textit{solid
magenta circles}), and Bouwens et al.\ 2004a (\textit{solid red
triangle}) are also shown.  The uncertainty expected from large scale
structure (cosmic variance) is $\pm$20\% for many of the lower
redshift points (e.g., Somerville et al.\ 2004) and $\pm$50\% for the
$z\sim7.5$ point.  The top horizontal axis provides the corresponding
age of the universe.  Lower Left Panel: The surface density vs. total
magnitude of the observed $z_{850}$-dropouts (\textit{histogram}) and
that predicted from a $(1+z)^{-1}$ size scaling of our GOODS
$B$-dropout sample (B04) (\textit{red}, see \S2e).  Lower Right Panel:
The expected redshift distribution for $z_{850}$-dropouts derived from
these same simulations.  These results suggest a modest to significant
decline in the star formation rate density (uncorrected for
extinction: see Thompson et al.\ 2004b for the extinction-corrected
star formation history).}
\end{figure}

\acknowledgements

We are appreciative to Andy Bunker, Dave Golimowski, Sandy Leggett,
Piero Madau, and Daniel Schaerer for useful conversations, Adam
Burgasser for important estimates of T dwarf surface densities, Sune
Toft for help with the PSFs, and our referee Haojing Yan for comments
which significantly improved this manuscript.  This research was
supported under NASA grant HST-GO09803.05-A and NAG5-7697.

\newpage
\begin{deluxetable}{lccccccccc}
\tablewidth{480pt} \tabletypesize{\scriptsize}
\tablecaption{$z\sim7-8.5$ Sample.\tablenotemark{a}} \tablehead{
\colhead{Object ID} & \colhead{Right Ascension} &
\colhead{Declination} & \colhead{$H_{160,Cor}$} &
\colhead{$H_{160,Ap1}$} & \colhead{$H_{160,Ap2}$} & \colhead{$z - J$}
& \colhead{$J - H$} & \colhead{S/G} & \colhead{$r_{hl}$(\arcs)}}
\startdata
UDF-825-950 & 03:32:39.538 & -27:47:17.41 & $26.1\pm0.3$ & $27.3\pm0.2$ & $26.7\pm0.2$ &
$>$2.1 & 0.5$\pm0.3$ & 0.08 & 0.39\\

UDF-491-880\tablenotemark{\dagger} & 03:32:40.941 & -27:47:41.83 & $26.6\pm0.3$ &
$27.6\pm0.3$ & $26.9\pm0.2$ & $>$2.3 & 0.1$\pm0.3$ & 0.03 & 0.34\\
UDF-387-1125 & 03:32:42.565 & -27:47:31.42 & $26.6\pm0.3$ &
$27.4\pm0.2$ & $27.0\pm0.2$ & 1.4$\pm$0.4 & 0.8$\pm0.3$ & 0.68 & 0.28\\
UDF-983-964 & 03:32:38.794 & -27:47:07.14 & $27.1\pm0.3$ & $27.8\pm0.3$ & $27.1\pm0.2$ &
$>$2.1 & 0.0$\pm0.3$ & 0.48 & 0.27\\
UDF-818-886\tablenotemark{\dagger} & 03:32:39.292 & -27:47:22.12 & $27.1\pm0.3$
& $27.6\pm0.3$ & $27.3\pm0.3$ & $>$2.0 & 0.2$\pm0.3$ & 0.84 & 0.23\\
UDF-640-1417\tablenotemark{*} & 03:32:42.562 & -27:46:56.58 &
$26.0\pm0.3$ & $27.1\pm0.2$ & $26.5\pm0.2$ & 1.1$\pm$0.2 & 0.7$\pm0.2$ & 0.11 & 0.37\\
\enddata

\tablenotetext{a}{All magnitudes are AB magnitudes.  Right ascension
and declination use the J2000 equinox.  Errors are $1\sigma$.  Limits
on $z_{850}-J_{110}$ colors are 2$\sigma$.  ``S/G'' denotes the
SExtractor stellarity parameter, for which 0 indicates an extended
object, and 1 a point source.  ``Cor'' refers to a total magnitude
estimated using the Kron system (see \S2), ``Ap1'' refers to a
0.6\arcsec-diameter aperture magnitude, and ``Ap2'' refers to a
1.0\arcsec-diameter aperture magnitude.  $z-J$ and $J-H$ colors were
estimated in a Kron aperture with Kron factor equal to 1.2 (similar to
the $0.6''$-diameter apertures used for ``Ap1'').}
\tablenotetext{\dagger}{These candidates are not found (UDF-818-886)
or seen to lower significance (UDF-491-880) in an independent
reduction of the NICMOS field kindly provided to us by Robberto et
al.\ (2004).}
\tablenotetext{*}{This object was very close to meeting
our selection criteria and could be a reddened starburst at
$z\sim6.5$.  Another possibility is that of a dusty/evolved galaxy at
$z\sim1.6$.  This object was also found by Yan \& Windhorst (2004).}
\end{deluxetable}
\end{document}